\documentclass[10pt,a4paper,twocolumn,english,prl,aps,showpacs,floatfix,groupedaddress,superscriptaddress]{revtex4-1}

\usepackage{graphicx}
\usepackage{epsfig}
\usepackage[english]{babel}
\usepackage{amsmath}
\usepackage{amssymb}
\usepackage{amsfonts}
\usepackage{bm}
\usepackage{bbm}
\usepackage{fixmath}
\usepackage{color,array}
\usepackage{dsfont}

\newcommand{\be}{\begin{equation}}
\newcommand{\ee}{\end{equation}}
\newcommand{\bes}{\begin{subequations}}
\newcommand{\ees}{\end{subequations}}

\newcommand{\ave}{_\text{ave}}
\newcommand{\rel}{_\text{rel}}

\newcommand{\ket}[1]{|#1\rangle}
\newcommand{\bra}[1]{\langle#1|}

\newcommand{\im}{\text{Im}}

\begin{document}
\title{Positivity of the spectral densities of retarded Floquet Green functions}

\author{G\"otz S.\ Uhrig}
\email{goetz.uhrig@tu-dortmund.de}
\affiliation{Lehrstuhl f\"{u}r Theoretische Physik I, Technische Universit\"at Dortmund, 
Otto-Hahn Stra\ss{}e 4, 44221 Dortmund, Germany}

\author{Mona H.\ Kalthoff}
\email{mona.kalthoff@tu-dortmund.de}
\affiliation{Lehrstuhl f\"{u}r Theoretische Physik I, Technische Universit\"at Dortmund, 
Otto-Hahn Stra\ss{}e 4, 44221 Dortmund, Germany}

\author{James K.\ Freericks}
\email{freericks@physics.georgetown.edu}
\affiliation{Department of Physics, Georgetown University, 37th St.\ and O St., 
NW, Washington, D.C.\ 20057, USA}

\date{August 30, 2018}

\begin{abstract}
Periodically driven nonequilibrium many-body systems are interesting because they have a 
quasi-energy spectra, which can be tailored
by controlling the external driving fields. We derive the general spectral representation
of retarded Green functions in the Floquet regime, thereby generalizing the well-known Lehmann
representation from equilibrium many-body physics. The derived spectral Floquet
representation allows us to prove the nonnegativity of spectral
densities and to determine exact spectral sum rules, which can be employed to 
benchmark the accuracy of approximations to the exact Floquet many-body Green functions.
\end{abstract}

\maketitle

% Intro: why interesting, non-equilibrium, quasi-stationarity, Floquet
% non-negativity: examples, why?

Nonequilibrium many-body physics is a vibrant field,
both from the experimental and from the theoretical side. Largely, this has been
triggered by the ease with which one can tune and manipulate the time dependence of systems of ultracold
atoms in optical lattices \cite{bloch08,essli10}. But there also have been significant advances in solid state systems, which employ ultrafast
pump-probe techniques to study electrons on femto-second timescales \cite{axt04,perfe06,kiril10}. 

Periodically driven many-body systems are simpler than general nonequilibrium systems, because the Hamiltonian repetitively cycles through the same functional form again and again. Conceptually, Floquet theory 
for periodic linear differential equations (and also periodic Hamiltonians)
\cite{floqu1883,grifo97,hangg98} is a powerful tool to treat these periodically
modulated quantum systems. One fundamental development is the notion of
Floquet design, i.e., the possibility to engineer quantum systems with certain
desired properties, e.g., with topological phases \cite{lindn12},
by properly selecting the external drive, see, e.g., Refs.\ 
\onlinecite{sente15,menti15,puvia16,kenne18}.  An important
issue for experimentally realizing such systems is how long the system has to be driven to display Floquet-like behavior
\cite{kalth18}?  It turns out that the drive time need not be so long for many of these systems, as has been experimentally demonstrated with topological insulators \cite{wang13b}.

In spite of the large interest in periodically driven many-body systems, rigorous
statements about the properties of measurable and computable quantities in the Floquet regime 
are scarce. As a relevant example, we draw the reader's attention to the fact that
the fermionic spectral density is nonnegative in equilibrium, allowing its interpretation as a probabilitistic density of states. But in Floquet systems, there is no \textit{a priori} guarantee that a spectral density will be nonnegative. As a result, others have employed weighted sums over various elements of the 
response functions in the Floquet representation \cite{frank13,puvia16,qin17b,qin18}, which
have turned out to be nonnegative. But to our knowledge, no proof of nonnegativity has been offered. This is a nontrivial issue, as the standard approach to constructing spectral funtions, which involves using Wigner coordinates of average and relative time, and Fourier transforming the relative time to a frequency, produces
spectral functions $A(\omega,t\ave)$ that usually
display negative values. However, they become nonnegative after further averaging over $t\ave$ 
\cite{potot14,gensk17,kalth18}. For noninteracting single-band models, 
 analytical proofs do exist that show how averaging over one period $T$ guarantees
nonnegative spectral densities \cite{tsuji08,kalth18}. Nevertheless, negative densities
are sometimes seen for interacting systems \cite{tsuji08}, so far without explanation. This illustrates
the need for tangible analytic results which hold also in presence of interactions.

We solve this problem by deriving
a spectral representation for retarded Green functions in the quasi-stationary Floquet regime.
This spectral representation generalizes the well-established Lehmann representation
of equilibrium quantum mechanics. Like the latter, our generalization allows one to
derive rigorous general conclusions, e.g., on the nonnegativity of spectral
functions and on their sum rules. For this reason, the derived results will guide
many future studies in the field.

%What is considered: H and quantities
We consider a closed quantum-mechanical system described by
the time-dependent periodic Hamiltonian
\be
\label{eq:periodic}
\mathcal{H}(t)=\mathcal{H}(t+T) \quad\forall\; t,
\ee
where $T$ is the temporal period. 
Hence, any linear-response function, e.g., a fermionic or bosonic progagator,
generically depends on two times $t_1$ and $t_2$ in a nontrivial way. In other words, the
relative time $t\rel$ alone is not sufficient, in contrast to time-invariant systems
with constant Hamiltonians $\partial_t \mathcal{H}=0$, where only the relative time dependence enters. 

Wigner's prescription for the relative and average times is given by
\bes
\begin{align}
t\rel &:= t_1- t_2
\\
t\ave &:= (t_1+ t_2)/2.
\end{align}
\ees

Kubo's formalism tells us that a retarded Green function satisfies
\be
\label{eq:green}
G(t_1,t_2) := -i\left\langle\left[ c(t_1),c^\dag(t_2)\right]_\pm\right\rangle 
\Theta(t_1-t_2),
\ee
where $c$ can be any, possibly composite, fermionic or bosonic operator,
e.g., a fermionic annihilation operator in position space or in momentum space
or a Hubbard operator.
If it is overall fermionic (that means odd in the number of
fermionic creation or annihilation operators), then the $+$ sign applies in the anticommutator $[\cdot]_+$;
if it is overall bosonic, then the $-$ sign applies in the commutator $[\cdot]_-$.
We re-express $G(t_1,t_2)$ in terms of $t\rel$ and $t\ave$ and transform
the dependence on $t\rel$ to frequency space
\begin{align}
\label{eq:fourier-general}
& G(\omega,t\ave) \ := 
\\
\nonumber
&\lim_{\delta\to0^+} \int_0^\infty e^{i(\omega+i\delta) t\rel}
G(t\ave+{t\rel/2},t\ave-{t\rel/2})dt\rel .
\end{align}
We are interested in the spectral function $A(\omega,t\ave)$, 
which we define by the negative imaginary part of the retarded Green
function $A(\omega,t\ave) := -\text{Im} G(\omega,t\ave)/\pi$ as usual.
We will derive a spectral representation for the Wigner representation \cite{wigne32,tsuji08}
of this quantity below. For clarity, we deal with the greater Green function 
$G^>(t_1,t_2) := -i\left\langle c(t_1) c^\dag(t_2)\right\rangle$
\cite{turko05}  in the explicit calculations; the expressions for the 
lesser Green function $G^<(t_1,t_2) := \pm i\left\langle c^\dag(t_2)  c(t_1)\right\rangle$ 
(the upper sign refers to fermions) are analogous. Finally, we are interested in the retarded Green function $G(t_1,t_2)=(G^>(t_1,t_2) - G^<(t_1,t_2))\Theta(t_1-t_2)$ \cite{stefa13} in order to define the spectral densities. Note that we expect there to be stronger average time dependence to the lesser and greater Green functions (due to heating effects) than to the retarded Green function, since the latter is determined primarily by the quantum states and not by how those states are occupied.

% Lehmann representation
For comparison, it is useful to recall the Lehmann representation for a time-independent Hamiltonian $\mathcal{H}$. 
Let us assume that $\{\ket{m}\}$ is an eigenbasis of $\mathcal{H}$ 
with eigenvalues $\epsilon_m$ and that the system is
in the state $\ket{m}$ with probability $p_m\ge 0$ and $\sum_m p_m = 1$. 
We do not necessarily require a thermal distribution $p_m\propto \exp(-\beta\epsilon_m)$ for these probabilities, but we do require monotonicity for the bosonic case, where $p_m\ge p_n$ if $\epsilon_m\le \epsilon_n$.
Then the greater Green function can be expressed as
\be
\label{eq:green-constant}
G^>(t_1,t_2) = -i\sum_{m,n} p_m |\bra{m}c \ket{n}|^2 e^{i(\epsilon_m-\epsilon_n)t\rel}.
\ee
The Fourier transform of Eq.~\eqref{eq:fourier-general} of the retarded Green function
leads to
\be
\label{eq:spectral-constant}
A(\omega) = \sum_{m,n} (p_m\pm p_n) |\bra{m}c \ket{n}|^2 \delta(\omega-(\epsilon_n-\epsilon_m)),
\ee
which does not depend on $t\ave$ because $\mathcal{H}$ is time-independent.
Recall that $A(\omega)$ is strictly nonnegative in the fermionic case and
in the bosonic case for $\omega\ge0$ since $\epsilon_n \ge \epsilon_m$ implies
$p_n \le p_m$ (due to our monotonicity requirement); for bosons and $\omega\le0$, $A(\omega)$ is nonpositive.

% Floquet theory, absolute basics, details in appendix
For periodic Hamiltonians, the Schr\"odinger equation
$i\partial_t \ket{\psi(t)} = \mathcal{H}(t) \ket{\psi(t)}$ 
is generally solved by the linear superposition of special solutions of the form
\cite{grifo97,hangg98}
\be
\label{eq:floquet-ansatz}
\ket{\psi_m(t)} = \exp(-i\epsilon_m t)\ket{m,t} \quad \forall \; m\in\mathbb{N}_0
\ee
where the Floquet states $\ket{m,t}$ are periodic in time with period $T$
$\ket{m,t+T} = \ket{m,t}$. This ansatz strongly reminds of the 
Bloch theorem transferred to time.
At any given instant $t$, the states $\ket{m,t}$ form a complete, orthonormal
basis. Hence, the unitary time evolution $U(t_1,t_2)$ from time $t_2$ to time $t_1$ 
is expressed by
\be
\label{eq:U}
U(t_1,t_2) = \sum_{m=0}^\infty \exp(-i\epsilon_m (t_1-t_2)) \, \ket{m,t_1} \bra{m,t_2}
\ee
which we will use next. These properties are derived in the Supplementary Material
\cite{supplement}.

% Calculation for Wigner representation
% interpretation
Now we derive the spectral representation in terms of the
Floquet states $\ket{m,t}$. Using $U(t_1,t_2)$ and assuming that the system was at 
some time $t_0$  in the Floquet state $\ket{m,t_0}$ with probability $p_m$,
 we obtain for the greater Green function
\begin{align}
&G^>(t_1,t_2) = 
\\
\nonumber
&\ -i\sum_{m=0}^\infty p_m \bra{m,t_0} U(t_0,t_1)\, c\, U(t_1,t_2)
\,c^\dag\, U(t_2,t_0) \ket{m,t_0}.
\end{align}
Inserting the result in Eq.~\eqref{eq:U} yields
\begin{align}
G^>(t_1,t_2) &= -i\sum_{m,n =0}^\infty
p_m e^{-i\epsilon_m(t_0-t_1)} \bra{m,t_1} c \ket{n,t_1}\cdot
\nonumber
\\
& e^{-i\epsilon_n(t_1-t_2)}
\bra{n,t_2} c^\dag \ket{m,t_2} e^{-i\epsilon_m(t_2-t_0)}.
\label{eq:lesser1}
\end{align}
The dependence on $t_0$ cancels out so that one may choose any
appropriate instant. We define the $T$-periodic functions
\be
\Phi_{m,n}(t) := \bra{m,t} c \ket{n,t}
\label{eq:Phi_def}
\ee
so that $\Phi_{m,n}^*(t) = \bra{n,t} c^\dag \ket{m,t}$,
and we can express Eq.~\eqref{eq:lesser1} by
\be
G^>(t_1,t_2) =-i \sum_{m,n =0}^\infty
p_m e^{i(\epsilon_m-\epsilon_n)t\rel} \Phi_{m,n}(t_1)
\Phi^*_{m,n}(t_2).
\label{eq:lesser_Greens_t1t2}
\ee
This result strongly resembles Eq.~\eqref{eq:green-constant},
but cannot be Fourier transformed directly due to the
time dependence of the functions  $\Phi_{m,n}(t)$. But the latter can 
be represented by a \textit{Fourier series} due to its time-periodicity
\be
\Phi_{m,n}(t)=\sum_{\alpha\in\mathbb{Z}} f^{(\alpha)}_{m,n} e^{i\alpha\Omega t},
\ee
where $\Omega=2\pi/T$. Inserting this expression (and its complex
conjugate) into the Wigner representation of the modified greater Green function yields
\be
\tilde G_\ell^>(\omega) := \int_0^\infty dt\rel 
e^{i\omega t\rel} \frac{1}{T} \int_{-\frac{T}{2}}^{\frac{T}{2}}
dt\ave
e^{il\Omega t\ave} \tilde G^> (t_1,t_2).
\ee
Here, the modified greater Green function has an additional $\Theta(t_{\text{rel}})$ multiplied in to allow us to produce the retarded Green function. The two integrals can be done, and yield
\bes
\be
\label{eq:Wigner_rep_G}
G_\ell^>(\omega) =
\sum_{m,n =0}^\infty p_m \sum_{\alpha\in\mathbb{Z}}
f^{(\alpha)}_{m,n} f^{(\alpha+l)*}_{m,n}\left\lbrace \frac{P}{\Delta\epsilon}
-i\pi\delta(\Delta\epsilon)\right\rbrace
\ee
where $P$ stands for the principal value of the pole and
\be
\Delta\epsilon := \omega-(\epsilon_n-\epsilon_m) + \left(\alpha+{\ell/2}\right)\Omega .
\ee
\ees
If we combine
this result with the analogous result for the similarly modified lesser Green function $\tilde G^<$ one 
obtains the Fourier coefficients of the Fourier series of the retarded spectral function 
$A(\omega,t\ave)=\sum_{\ell\in\mathbb{Z}}A_\ell(\omega)\exp(-i\ell\Omega t\ave)$
\bes
\label{eq:specrepfloq}
\begin{align}
A_\ell(\omega) &= -\frac{1}{\pi}\im G_\ell(\omega)
\\
&= \sum_{m,n =0}^\infty (p_m\pm p_n) \sum_{\alpha\in\mathbb{Z}}
f^{(\alpha)}_{m,n} f^{(\alpha+\ell)*}_{m,n} \delta(\Delta\epsilon),
\end{align}
\ees
where $+$ refers to fermionic operators and $-$ to bosonic ones.
This equation yields the general spectral representation of Floquet 
response functions; it generalizes the Lehmann representation in
equilibrium and is the key result of our Letter.

%%%

% Discussion
What can be deduced from Eq.~\eqref{eq:specrepfloq}? For $\ell\ne 0$ we do not see
any possibility for a general conclusion on positivity or reality of the spectral function.
But for $\ell=0$,  it is obvious that 
\bes
\label{eq:nonnegative}
\begin{align}
A_0(\omega) \!&=\!\!\!\sum_{m,n =0}^\infty (p_m+ p_n) \sum_{\alpha\in\mathbb{Z}}
\left|f^{(\alpha)}_{m,n}\right|^2 \delta(\omega-\epsilon_n+\epsilon_m + \alpha\Omega)
\\ 
&\ge0
\end{align}
\ees
in the fermionic case, i.e., $A_0(\omega)$ is nonnegative and can be interpreted as a
density-of-states just like in equilibrium. This conclusion is closely related to Bochner's
theorem \cite{bochn32}. Note that no general conclusion is possible in the
bosonic case since the interplay of the factor $(p_m-p_n)$ and the shift $\alpha\Omega$
can be complicated. We stress that the case $\ell=0$ corresponds precisely to the
average of $A(\omega,t\ave)$ over one period of $t\ave$ as we used previously 
\cite{kalth18} to reach physically meaningful results in the noninteracting case. Other authors have also averaged over one period to avoid negative spectral
densities \cite{potot14,gensk17}, but without explaining why the results must be nonnegative. The above derivation puts this averaging procedure 
on a firm mathematical basis.

%%% sum rules
Sum rules are another useful spin-off from spectral representations. Using
Eq.~\eqref{eq:specrepfloq},  we consider the zeroth-moment sum rule $S$ and obtain 
\bes
\begin{align}
S&:=\int_{-\infty}^\infty A_0(\omega) d\omega
\\ 
&= \sum_{m,n =0}^\infty (p_m\pm p_n) \sum_{\alpha\in\mathbb{Z}}
\left|f^{(\alpha)}_{m,n}\right|^2
\\
&= \frac{1}{T}\sum_{m,n =0}^\infty (p_m\pm p_n) \int_{t}^{t+T}|\Phi_{m,n}(t')|^2 dt'
\end{align}
\ees
where the last step results from Parseval's identity. Re-inserting
the definition from Eq.~\eqref{eq:Phi_def} for $\Phi_{m,n}(t)$ and using the completeness
relation \cite{supplement}
$\mathbbm{1}=\sum_n\ket{n,t}\bra{n,t}$, we arrive at the general sum rule
\be
\label{eq:sumrule}
S=\frac{1}{T}\sum_m p_m \int_{t}^{t+T} \bra{m,t'} [c,c^\dag]_\pm \ket{m,t'} dt'
\ee
which is consistent with the value of $G(t+0,t)$ in \eqref{eq:green}
averaged over one period $T$.
While in equilibrium, the sum rule is given by the expectation value of 
the (anti)commutator for (fermionic) bosonic operators, it is given
by the temporal average in the Floquet regime. Hence, we find tangible
evidence that the equivalent of a constant expectation value or a constant spectral
density at equilibrium is the temporal average of such an expectation value or
of such a spectral function, respectively.
The sum rules for higher moments of the spectral densities are commutators of products 
of operators in time so that they become convolutions after Fourier 
transformations in Floquet representation.
Examples of such sum rules are given in the Supplementary Material \cite{supplement}.

The sum rule in Eq.~\eqref{eq:sumrule} is particularly meaningful if we consider
fermionic or bosonic single-particle progagators, i.e., $c$ is a single-particle
annihilation operator and $c^\dag$ the corresponding  creation operator.
Then, every expectation value on the right hand side equals unity and so does
the temporal average and the weighted sum. Hence, the sum rule is indeed rigorously
the same as in equilibrium for the averaged spectral functions. We then conclude
that a fermionic spectral density in the Floquet regime can be interpreted to be
a density-of-states similar to what happens in equilibrium. This has been
used already in many numerical studies, see for instance Refs.\ \onlinecite{tsuji08,qin17b,qin18}.

Finally, we transform from the Wigner representation to the often employed 
equivalent Floquet representation. They are related by
\be
\label{eq:Floquet}
G_{\ell j}\left(\omega\right) :=
G_{\ell-j}\left(\omega+\Omega (\ell+j)/2\right),
\ee
where $\ell,j\in\mathbb{Z}$ according to Tsuji et al.\ \cite{tsuji08}. 
It is obvious that the Floquet representation does not contain more 
information than the Wigner representation. Indeed, the Floquet representation
is redundant unless one restricts its argument $\omega$ to the interval
$(-\Omega/2,\Omega/2]$ \cite{tsuji08}, but this restriction is not needed otherwise.
Equation \eqref{eq:Floquet} implies that the physically meaningful time-averaged
Green functions appearing in the Wigner representation at index zero occur in
the Floquet representation along the diagonal, i.e., for $\ell=j$. One has
$G_{\ell \ell}\left(\omega\right) = G_{0}\left(\omega+\ell\Omega\right)$
where different indices $\ell$ correspond to different shifts relative
to $G_{00}$. This Green function and the spectral density $A_0(\omega)$
stemming from its imaginary part are generically used
\cite{tsuji08,qin17b,qin18} because they behave like equilibrium spectral
densities. The negative spectral densities found in Ref.\ \onlinecite{tsuji08}
for the gauge-invariant Green function are due to the gauge-invariant transformation, which can no longer be proven to be nonnegative.

For completeness, we also provide the general expression for the
nondiagonal Floquet spectral functions $A_{\ell j}(\omega)=
-\im G_{\ell j}(\omega)/\pi$ which reads
\begin{align}
A_{\ell j}(\omega) &= \sum_{m,n =0}^\infty
\left(p_m\pm p_n\right)
\sum_{\alpha\in\mathbb{Z}}
f^{(\alpha-l)}_{m,n}
f^{(\alpha-j)*}_{m,n}\times 
\nonumber\\
&\quad \delta\left(\omega- (\epsilon_n-\epsilon_m)+\alpha\Omega\right).
\end{align}
This expression helps to understand why one obtains a positive spectral function
upon summing over all Floquet indices $\ell$ and $j$ as done in Ref.\ \onlinecite{frank13}.
Clearly
\bes
\begin{align}
&A_\Sigma(\omega):=\sum_{\ell,j\in\mathbb{Z}} A_{\ell j}(\omega)
\\
\label{eq:frank2}
&= \sum_{m,n =0}^\infty \left(p_m\pm p_n\right)
\left|F_{m,n}\right|^2 \sum_{\alpha\in\mathbb{Z}}
\delta\left(\omega- \epsilon_n+\epsilon_m+\alpha\Omega\right)
\end{align}
\ees
which also yields a nonnegative spectral density with
\be
F_{m,n} :=\sum_{\ell\in\mathbb{Z}} f^{(\ell)}_{m,n} =\Phi_{m,n}(t=0).
\ee
Note that no dependence on $\alpha$ remains except a shift by $\alpha\Omega$.
Thus, the sum over $\alpha$ on the right hand side of Eq.~\eqref{eq:frank2}
implies a divergence. But if we fix $\alpha$ to one single value or normalize with respect
to the number of Floquet replicas considered for this purpose, one obtains a nice sum rule again
\bes
\begin{align}
\sum_{m,n =0}^\infty \left(p_m\pm p_n\right) \left|F_{m,n}\right|^2 &
=\sum_m p_m \bra{m,0} [c,c^\dag]_\pm \ket{m,0}
\\
&= \left\langle [c,c^\dag]_\pm \right\rangle\Big|_{t=0}
\\
&= 1,
\end{align}
\ees
where the last equation holds for $c$ a fermionic or bosonic single-particle 
annihilation operator.

Summarizing, we considered a broad range of nonequilibrium systems which are
in the Floquet regime, i.e., they are given in a mixture of quasi-stationary Floquet states
which solve the time-dependent Schr\"odinger equation. For this 
setting, we rigorously established a generalization of the Lehmann representation.
The spectral representation of two-time Floquet response functions include the cases of fermionic 
and bosonic single-particle propagators. We clarified the relation to  
the Wigner representation, which exploits the periodicity in the average time
of the two times and to the Floquet representation. 

Our results show precisely when fermionic spectral functions
must be nonnegative and can be interpreted as densities-of-states. 
We also established some exact sum rules. 

As an outlook, we think that more information on the mathematical
properties of the self-energy in the Floquet regime is also desirable.
In equilibrium, for instance, one deduces from the Dyson equation that
the imaginary part of the self-energy is also nonnegative. Does 
a similar result holds in the Floquet regime?
One might conjecture that the self-energy averaged over $t\ave$ should also behave as
in equilibrium. But the Floquet Dyson equation is too complicated and does not appear to permit
one to establish this fact.

\textit{Acknowledgments:}
We acknowledge useful discussions with Joachim Stolze.
This work was supported by the Deutsche Forschungsgemeinschaft in Project No.
UH 90-13/1 (GSU) and by the Department of Energy, Office of Basic Energy
Sciences, Division of Materials Sciences and Engineering
(DMSE) under Contract No. DE-FG02-08ER46542 (JKF); JKF also 
acknowledges financial support from the McDevitt Bequest at Georgetown.
MHK acknowledges the financial support by the Studienstiftung des Deutschen Volkes. 

\bibliographystyle{apsrev}
%\bibliography{../../bibinput/liter10}
%\bibliography{liter10}

\section{Supplemental Material}
\section{Principles of Floquet theory}
\label{suppl:floquet}

Here we formally derive the properties of Floquet theory
used in the main text. We consider the unitary time evolution operator
$U(t_1,t_2)$ and in particular $U(T,0)$.
Any unitary operator such as $U(T,0)$ has an orthonormal eigenbasis
$\{ \ket{\psi_m} \}$ satisfying
\be
\label{eq:eigenbasis}
U(T,0) \ket{\psi_m} = \lambda_m  \ket{\psi_m}.
\ee
Since $U(T,0)$ is unitary, the absolute value of $\lambda_m$
is unity, so it can be written as
\be
\lambda_m = \exp(-i\epsilon_m T)
\ee
where the quasi-energy $\epsilon_m$ is uniquely defined only
if it is restricted to the interval $\epsilon_m\in (-\pi/T,\pi/T]$.
This is the temporal equivalent of a Brillouin zone.

Next, we take the states $\ket{\psi_m}$ as initial states, i.e.,
$\ket{\psi_m (t=0)}=\ket{\psi_m}$ for
a time-evolution according to the Schr\"odinger equation 
\be
\label{eq:schrodinger}
i\partial_t \ket{\psi(t)} = \mathcal{H}(t) \ket{\psi(t)}.
\ee
We emphasize that the orthonormality and the completeness
persist in the course of the time evolution because it is unitary
\bes
\label{eq:on-psim}
\begin{align}
\bra{\psi_m(t)}\psi_n(t)\rangle &= \delta_{mn}
\\
\label{eq:on-psimb}
\sum_m \ket{\psi_m(t)} \bra{\psi_m(t)} &= \mathbbm{1}.
\end{align}
\ees
But these relations only hold if the time arguments in bra and ket
are the same. Since the states $\ket{\psi_m(t)}$ are solutions of the Schr\"odinger
equation
\be
U(t_1,t_2)  \ket{\psi_m(t_2)} = \ket{\psi_m(t_1)}
\ee
holds by definition for all times $t_1$ and $t_2$. 
Thus, the unitary time evolution is given by
\be
\label{eq:time-evolution}
U(t_1,t_2) = \sum_m \ket{\psi_m(t_1)} \bra{\psi_m(t_2)}.
\ee
One can verify that this solves the Schr\"odinger equation
\bes
\begin{align}
i\partial_{t_1} U(t_1,t_2) &= i\partial_{t_1} U(t_1,t_2)
 \sum_m \ket{\psi_m(t_2)} \bra{\psi_m(t_2)}
\\
&=  \sum_m i\partial_{t_1} \ket{\psi_m(t_1)} \bra{\psi_m(t_2)}
\\
&=  \sum_m \mathcal{H}(t_1) \ket{\psi_m(t_1)} \bra{\psi_m(t_2)}
\\
&=  \mathcal{H}(t_1) U(t_1,t_2).
\end{align}
\ees
where we used that the states $\ket{\psi_m(t_1)}$ fulfill
the Schr\"odinger equation in Eq.~\eqref{eq:schrodinger}. The initial condition
\be
U(t_2,t_2) = \mathbbm{1}
\ee
is fulfilled due to the completeness in Eq.~\eqref{eq:on-psimb}
of the states $\{ \ket{\psi_m(t)}\}$.

By construction [see Eq.~\eqref{eq:eigenbasis}], the property
\be
\ket{\psi_m(T)} = U(T,0)\ket{\psi_m} = \exp(-i\epsilon_m T) \ket{\psi_m}
\ee
holds. More generally, quasi-periodicity holds
\bes
\begin{align}
\ket{\psi_m(t+T)} &=U(t+T,T) \ket{\psi_m(T)}
\\
\label{eq:2nd_line}
&= U(t,0) \exp(-i\epsilon_m T) \ket{\psi_m(0)}
\end{align}
\ees
resulting from the periodicity of the unitary time evolution,
which in turn is implied by the periodicity of the Hamiltonian and
of Eq.~\eqref{eq:eigenbasis}. Combining the unitary operator 
and the ket in Eq.~\eqref{eq:2nd_line} yields
\be
\label{eq:quasi-period}
\ket{\psi_m(t+T)} = \exp(-i\epsilon_m T) \ket{\psi_m(t)}
\ee
which confirms that $\ket{\psi_m(t)}$ is periodic
\emph{up to the factor} $\exp(-i\epsilon_m T)$.
This is what is conventionally regarded as the Floquet theorem.

Finally, we define the states used in Eq.~(7) via
\be
\label{eq:define-m}
\ket{m,t} := \exp(i\epsilon_m t) \ket{\psi_m(t)}.
\ee
Clearly, these states are periodic, inheriting this property from
the quasi-periodicity in Eq.~\eqref{eq:quasi-period} of $\ket{\psi_m(t)}$.
In addition, they form an orthonormal basis
\bes
\label{eq:on-m}
\begin{align}
\bra{m,t} n,t\rangle &= \delta_{mn}
\\
\sum_m \ket{m,t} \bra{m,t} &= \mathbbm{1}
\end{align}
\ees
which results again from the orthonormality in Eq.~\eqref{eq:on-psim} of the states $\ket{\psi_m(t)}$.

The representation of the time evolution operator in Eq.~\eqref{eq:time-evolution}
can be expressed in terms of the states $\ket{m,t}$
as well
\be
U(t_1,t_2) = \sum_m \exp(-i\epsilon_m(t_1-t_2))\, \ket{m,t_1} \bra{m,t_2}
\ee
which confirms Eq.~(8). Thereby, all properties used in the 
main text are derived.

\section{Sum rules for higher moments of the spectral densities in the Hubbard model}

We already discussed the zeroth moment sum rule in Eq.~(19), which is valid for any given Hamiltonian.
To analyze higher spectral moment sum rules, we have to specify the underlying model, as the sum rules depend on the particular form of the
Hamiltonian. Here we will present results for the Hubbard Hamiltonian, which is one of the simplest models to describe 
electron-electron interactions. Furthermore, it is a model for which the sum rules are well-known
\cite{turko08}.
The Hubbard-Hamiltonian is given by 
\begin{align}
\mathcal{H}_\mathrm{H}\left(t\right)
=&
-\sum_{ij\sigma}
t_{ij}\left(t\right)c_{i\sigma}^\dagger c_{j\sigma}^{\phantom\dagger}
+
\sum_i
U_i\left(t\right)
n_{i\downarrow}n_{i\uparrow}
\\&
-
\sum_i 
\mu_i\left(t\right)
\left(
n_{i\downarrow}+n_{i\uparrow}
\right)\,,\nonumber
\end{align} 
where $t_{ij}\left(t\right)$ is the time-dependent Hermitian electron hopping matrix, $U_i\left(t\right)$ is the time-dependent on-site Hubbard repulsion,
and $\mu_i\left(t\right)$ is a time-dependent  local site energy. To simplify the formulas, we introduce the notation $\left[\tilde{O}=\hat O\left(t\ave\right)\right]$
to indicate the operator (or function) is evaluated at the average time $t\ave$ after taking the limit $t\rel\rightarrow 0$. We assume that $\tilde{t}_{ij}$, 
$\tilde{U}_i$ and $\tilde{\mu}_i$ are $T$ periodic in $t\ave$ and therefore can be written as a Fourier series 
\begin{align}
\tilde{t}_{ij}=&
\sum_n t_{ij}^{n} \mathrm{exp}\left[in\frac{2\pi}{T}t\right]
\end{align}
(analogous for $\tilde{U}_i$ and $\tilde{\mu}_i$). The zeroth moment sum rule is given by 
$\mu_{ij\sigma}^{R0}\left(t\ave\right)=\delta_{ij}$, so integrating over one period
\begin{equation}
 \frac{1}{T}\int_x^{x+T}
\mu_{ij\sigma}^{R0}\left(t\ave\right)
\,\mathrm{d}t\ave
=
\delta_{ij}
\end{equation}
does not change the result. This is different for the first moment, which is given by
\begin{equation}
 \mu_{ij\sigma}^{R1}\left(t\ave\right)=
-\tilde{t}_{ij}-\delta_{ij}\tilde{\mu}_i+\delta_{ij}\tilde{U}_i\left\langle\tilde{n}_{i\tilde{\sigma}}\right\rangle\,,
\end{equation}
so the integration yields 
\begin{align}
\frac{1}{T}\int_t^{t+T}
\mu_{ij\sigma}^{R1}\left(t\ave\right)
\,\mathrm{d}t\ave
=&
t_{ij}^0
-
\delta_{ij}
\mu_i^0
\\&+
\delta_{ij}
\sum_m
U_i^m
\left\langle n_{i\bar{\sigma}}\right\rangle^{-m}\,.\nonumber
\end{align}
The second moment sum rule is given by 
\begin{eqnarray}
\mu_{ij\sigma}^{R2}\left(t\ave\right)
&=&
\sum_k
\tilde{t}_{ik}
\tilde{t}_{kj}
\\\nonumber
&&+
\tilde{t}_{ij}
\tilde{\mu}_i
+
\tilde{t}_{ij}
\tilde{\mu}_j
-
\tilde{t}_{ij}
\tilde{U}_i
\left\langle\tilde{n}_{i\bar{\sigma}}\right\rangle
-
\tilde{t}_{ij}
\tilde{U}_j
\left\langle\tilde{n}_{j\bar{\sigma}}\right\rangle
\\\nonumber
&&+
\delta_{ij}\left(
\tilde{\mu}_i^2
+
{\color{black}
\tilde{U}_i^2
\left\langle\tilde{n}_{i\bar{\sigma}}\right\rangle^2
}
-
2\tilde{\mu}_i\tilde{U}_i
\left\langle\tilde{n}_{i\bar{\sigma}}\right\rangle
\right)
\\\nonumber
&&+
\delta_{ij}\left(
\tilde{U}_i^2
\left\langle\tilde{n}_{i\bar{\sigma}}\right\rangle
-
{\color{black}
\tilde{U}_i^2
\left\langle\tilde{n}_{i\bar{\sigma}}\right\rangle^2}
\right)
\end{eqnarray}
which, when integrated over one period becomes
\begin{eqnarray}
&&\frac{1}{T}\int_x^{x+T}
\mu_{ij\sigma}^{R2}\left(t\ave\right)
\,\mathrm{d}t\ave
=
\sum_{k,n}
t_{ik}^n
t_{kj}^{-n}
\\\nonumber
&&\quad+
\sum_n\left(
t_{ij}^n
\mu_i^{-n}
+
t_{ij}^n
\mu_j^{-n}
\right)
\\\nonumber
&&\quad-
\sum_{nm}
\left(
t_{ij}^{n+m}
U_i^{-n}
\left\langle n_{i\bar{\sigma}}\right\rangle^{-m}
+
t_{ij}^{n+m}
U_j^{-n}
\left\langle n_{j\bar{\sigma}}\right\rangle^{-m}
\right)
\\\nonumber
&&\quad+
\delta_{ij}
\sum_n
{\color{black}{\left|\mu_i^n\right|^2}}
\\\nonumber
&&\quad
-
2\delta_{ij}
\sum_{mn}
\mu_i^{n+m}
{U}_i^{-n}
\left\langle n_{i\bar{\sigma}}\right\rangle^{-m}
\\\nonumber
&&\quad+
\delta_{ij}
\sum_{mn}
U_i^{n+m}
U_i^{-n}
\left\langle n_{i\bar{\sigma}}\right\rangle^{-m}\,.
\end{eqnarray}
It is obvious that the mixing of Floquet coefficients increases as we go to higher moments.

Finally we would like to discuss the zeroth moment of the 
self energy, given by 
\begin{eqnarray}
C_{ij\sigma}^{R0}\left(t\ave\right)
&=&
\delta_{ij}\left(
\tilde{U}_i^2
\left\langle\tilde{n}_{i\bar{\sigma}}\right\rangle
-
\tilde{U}_i^2
\left\langle\tilde{n}_{i\bar{\sigma}}\right\rangle^2
\right)\,.
\end{eqnarray}
Here the integration over one period yields 
\begin{eqnarray}
&&\frac{1}{T}\int_x^{x+T}\nonumber
C_{ij\sigma}^{R0}\left(t\ave\right)
\,\mathrm{d}t\ave
=
\delta_{ij}
\sum_{mn}
U_i^{n+m}
U_i^{-n}
\left\langle n_{i\bar{\sigma}}\right\rangle^{-m}\\
&&\qquad\qquad-
\delta_{ij}
\sum_{lmn}
U_i^{l+m+n}
U_i^{-l}
\left\langle n_{i\bar{\sigma}}\right\rangle^{-m}
\left\langle n_{i\bar{\sigma}}\right\rangle^{-l}\,,
\end{eqnarray}
so even for the lowest moment of the self energy, the Fourier coefficients of $\tilde{U}_i$ and $\tilde{n}_i$ mix.

\end{document}